\documentclass[12pt]{article}
\usepackage{amsmath}
\usepackage{amsfonts}   
\usepackage{amssymb}
\def\a{\alpha}

\def\r{\rho}
\def\s{\sigma}
\def\m{\mu}
\def\n{\nu}
\def\p{\partial}
\begin{document}
\date{}
\title{\textbf{Hawking Radiation due to Photon and Gravitino Tunneling}}
\author{{Bibhas Ranjan Majhi}\thanks{E-mail: bibhas@bose.res.in}\\
\\\textit{S.~N.~Bose National Centre for Basic Sciences,}
\\\textit{JD Block, Sector III, Salt Lake, Kolkata-700098, India}\\\\\\
{Saurav Samanta}\thanks{E-mail: srvsmnt@gmail.com}\\
\\\textit{Narasinha Dutt College,}
\\\textit{129, Belilious Road, Howrah-711101, India}}
\maketitle
                                                                                
\begin{quotation}
\noindent \normalsize
Applying the Hamilton--Jacobi method we investigate the tunneling of photon across the event horizon of a static spherically symmetric black hole. The necessity of the gauge condition on the photon field, to derive the semiclassical Hawking temperature, is explicitly shown. Also, the tunneling of photon and gravitino beyond this semiclassical approximation are presented separately. Quantum corrections of the action for both cases are found to be proportional to the semiclassical contribution. Modifications to the Hawking temperature and Bekenstein-Hawking area law are thereby obtained. Using this corrected temperature and Hawking's periodicity argument, the modified metric for the Schwarzschild black hole is given. This corrected version of the metric, upto $\hbar$ order is equivalent to the metric obtained by including one loop back reaction effect. Finally, the coefficient of the leading order correction of entropy is shown to be related to the trace anomaly. 
\end{quotation}

\section{Introduction}
Black holes are the solution of classical general relativity from which nothing can escape. In 1974--75 this understanding was changed completely when Hawking\cite{Hawking1,Hawking2} showed that due to quantum effects black holes can radiate energy and the resulting spectrum is purely thermal in nature. He also showed that the temperature of a black hole is directly proportional to its surface gravity.

To understand Hawking radiation in a physically intuitive manner, Parikh and Wilczek\cite{Wilczek} described it as a quantum tunneling effect through the horizon of a black hole. In this method one first calculates the tunneling amplitude by exponentiating the imaginary part of the action for outgoing mode, for the process of $s$- wave emission and then uses the principle of detailed balance to relate it with the Boltzmann factor. In the literature two different approaches are available to compute the imaginary part of the action that yields the Hawking temperature. In one method, trajectory of a radial null geodesic is considered--this was developed in \cite{Wilczek}. In the other method, Hamilton-Jacobi ansatz is used--this is an extension of the complex path analysis given in \cite{Paddy}. After the initial formulation of the theory it generated a lot of interest and till now it has been applied successfully to various types of black holes and space times \cite{Vanzo,Kim,Jiang,Bibhas,RB}. It was also noticed that in this approach there was a problem of factor $2$ in the expression of Hawking temperature. This was later solved in \cite{dog} by taking into account the temporal contribution to the quasi–classical amplitude. However, most of the studies in this context have been done for spinless scalar particles. Though there are few papers \cite{Chen,Majhifermi} on spin $\frac{1}{2}$ fermion tunneling, no analysis has been done for a spin one particle like photon. Discussion on the radiation of spin $\frac{3}{2}$ gravitino has been done recently \cite{Mann}, but that study incorporates only semiclassical approximation and does not consider quantum corrections.

   In the present paper, we shall study the tunneling of photon and gravitino from the horizon of a static spherically symmetric black hole by following the method previously elaborated by one of us \cite{Majhibeyond,Majhiback,Majhifermi,Debraj} and which has been applied later for various cases \cite{Modak}. This approach is basically the Hamilton--Jacobi method where quantum corrections to the usual semiclassical results are taken by considering all the terms in the expansion of the action. It was shown that all the higher order corrections are proportional to the semiclassical contribution. Though the values of these constants depend on the order of expansion, a general form was provided by simple dimensional argument. By calculating the ratio between outgoing and ingoing probability of a particle, Hawking temperature with quantum corrections was obtained. This eventually led to the entropy of the black hole in which the first order correction was logarithmic in nature.

     Here we employ the same method to first study the tunneling of photon. First we take the gauge fixed action in a static spherically symmetric space time background. Variation of this action with respect to the gauge field `$A_\mu$' gives the gauge fixed Maxwell equation. Substituting the standard ansatz for `$A_\mu$' and taking the semiclassical limit ({\it {i.e.}} $\hbar\rightarrow 0$) we obtain the usual semiclassical Hamilton-Jacobi equation. Solutions of this equation lead to the ingoing and outgoing probabilities of the gauge particle. Then applying the principle of detailed balance the usual Hawking temperature is identified.

    Since later we shall extend our analysis to higher order in $\hbar$, the above procedure is not convenient. So for simplicity we use Lorentz gauge condition separately.  Therefore we shall start with the $U(1)$ Maxwell action without any gauge fixed term. An arbitrary variation of $A_{\mu}$ in this action gives the standard Maxwell equation. Now substituting the previous ansatz for `$A_\mu$' and taking $\hbar\rightarrow 0$ limit, the Hamilton-Jacobi equation is obtained. This cannot be solved by the previous method because of the presence of different polarization vectors. To have a relation between these vectors we impose the Lorentz gauge condition. Substitution of the same ansatz for `$A_\mu$' in this gauge condition leads to another equation. Simultaneously solving these two equations we obtain the desired Hamilton-Jacobi equation which was derived directly from the gauge fixed action. 

     The photon tunneling beyond semiclassical approximation upto $\hbar$ order is also discussed here. In this case, the action and the polarization vectors are expanded in powers of $\hbar$. Then equating different powers of $\hbar$ on both sides of the Maxwell equation and the gauge condition we obtain a series of equations. These equations are simplified by using the previous equations in a recursive manner. Here we adopt only the second formalism where Maxwell equation and corresponding gauge condition are treated separately. Because in the other method simplification of $\hbar$ order equation by using $\hbar^0$ order equation is very difficult. This analysis again convinces the usefulness of gauge condition. After simplification we show that $\hbar$ order term of the action is proportional to the semiclassical contribution. This study for the photon field is completely new and has not been mentioned elsewhere.  

After obtaining the explicit form of the action, we calculate the wave function which is finally used to get the tunneling amplitude. We again apply the detailed balance principle to get the modified Hawking temperature. The result agrees upto $\hbar$ order with the conclusions previously obtained for the tunneling of scalar \cite{Majhibeyond,Majhiback} and Dirac particles \cite{Majhifermi} which confirms the robustness of the whole formalism. A point we want to mention here, is that in our analysis, the corrected tunneling amplitude is exactly the Boltzmann factor $e^{-\frac{\omega}{T_h}}$, where $T_h$ is the corrected Hawking temperature. In addition, there are also approaches \cite{KW} which lead to a different type of correction to the tunneling amplitude, that is essentially non-thermal in nature.

  We next study the gravitino tunneling beyond semiclassical approximation. For that we consider the massless Rarita-Schwinger equation \cite{Ra} in curved geometry and follow the same formalism. Though the final results for both photon and gravitino tunneling look similar upto $\hbar$ order, the difference comes from the correction parameter which is later shown to be dependent on the spin of the particle.

  By using Hawking's periodicity arguments for the temperature corrected upto order $\hbar$, we also give the corrections of the Schwarzschild metric in our paper. This is shown to be exactly equivalent to the result obtained in \cite{JM} by incorporating the one loop back reaction effect in the space time. Also the leading order correction term in the Bekenstein-Hawking area law is obtained, which is given as the logarithmic of the usual horizon area. Finally, application of the constant scale transformation in the metric coefficients reveals that the coefficient of this correction is related to trace anomaly. 

Before proceeding further, let us mention the organization of our paper. In the second section we study the tunneling of photon by two different methods in two separate subsections. In subsection 2.1 we consider the gauge fixed action for the photon field. In the next subsection, we consider the standard Maxwell action but impose the Lorentz gauge condition later to find the semiclassical black hole temperature. In the third section, the first order quantum effect to the photon tunneling is studied to find the modified Hawking temperature. Gravitino tunneling is analyzed in the next section. The discussion on the correction parameter is given in section 5 and the last section is for conclusions.

\section{Photon tunneling and Hawking temperature}
In this section we study the tunneling of photon to calculate the Hawking temperature of a black hole. This is done by following two methods in two subsections. In the first method we start from the gauge fixed action of Maxwell field in a curved spacetime background  and then find the action by using the Hamilton--Jacobi equation\cite{Paddy} to calculate the tunneling amplitude. Finally, this is equated with the Boltzmann factor to get the black hole temperature. In the other method we perform a similar analysis. However instead of the gauge fixed action, we take the standard photon field action and impose the Lorentz gauge condition later to obtain the tunneling amplitude. In both the analysis, we use the semiclasical approximation $\hbar\rightarrow 0$. 
\subsection{Method 1: Gauge fixed equation of motion}
Throughout this paper we shall consider the background space-time to be static and spherically symmetric in nature, {\it i.e.}
\begin{eqnarray}
ds^2=-f(r)dt^2+\frac{dr^2}{g(r)}+r^2d\Omega^2\label{ds2}
\end{eqnarray}
whose horizon $r=r_H$ is given by $f(r_H)=g(r_H)=0$.

The electromagnetic field in a gravitational background is described by the Lagrangian density
\begin{eqnarray}
\mathcal L=-\frac{1}{4}\sqrt{-g}F_{\m\n}F^{\m\n}\label{lagd}
\end{eqnarray}
where the field strength $F_{\m\n}$ is defined in terms of the gauge field $A_{\m}$ as,
\begin{eqnarray}
F_{\m\n}&=&\nabla_{\m}A_{\n}-\nabla_{\n}A_{\m}\\&=&\p_{\m}A_{\n}-\p_{\n}A_{\m}.\label{fmunu}
\end{eqnarray}
Under a local gauge transformation
\begin{eqnarray}
A_{\m}\rightarrow A'_{\m}=A_{\m}+\nabla_{\m}\Lambda,
\end{eqnarray}
the Lagrangian density (\ref{lagd}) is invariant. In order to quantize the theory, this symmetry is broken by adding a gauge fixing term
\begin{eqnarray}
{\mathcal L}_G=-\frac{1}{2}\xi^{-1}(\nabla_{\m}A^{\m})^2
\end{eqnarray}
to (\ref{lagd}) to get the following action
\begin{eqnarray}
S=\int {\textrm d}^4x(\mathcal L+{\mathcal L}_G)=-\int {\textrm d}^4x[\frac{1}{4}\sqrt{-g}F_{\m\n}F^{\m\n}+\frac{1}{2}\xi^{-1}(\nabla_{\m}A^{\m})^2].\label{77}
\end{eqnarray}
In this subsection we shall work with this action. Variation of the above action with respect to $A_{\m}$ gives the equation of motion
\begin{eqnarray}
\Box A_{\m}+R_{\m}^{\rho}A_{\rho}-(1-\xi^{-1})\nabla_{\m}(\nabla_{\n}A^{\n})=0.
\end{eqnarray}
Choosing $\xi=1$ (Feynman gauge) we get the simplified equation
\begin{eqnarray}
\Box A_{\m}+R_{\m}^{\rho}A_{\rho}=0.
\end{eqnarray}
Above equation is written explicitly in terms of the Christoffel connection and Ricci tensor as
\begin{eqnarray}
g^{\rho\sigma}\Big(\p_{\rho}\p_{\sigma}A_{\m}-2\Gamma^{\lambda}_{\sigma\m}\p_{\rho}A_{\lambda}-\Gamma^{\lambda}_{\r\s}\p_{\lambda}A_{\m}-\p_{\r}\Gamma^{\lambda}_{\sigma\m}A_{\lambda}+\Gamma^{\lambda}_{\r\s}\Gamma^{\a}_{\lambda\m}A_{\a}+\Gamma^{\lambda}_{\r\m}\Gamma^{\a}_{\s\lambda}A_{\a}\Big)+R_{\m}^{\rho}A_{\rho}=0.
\label{eom3}
\end{eqnarray}
This equation can be solved by using spherical harmonics technique as was done in \cite{dn} for the scalar field. But here we shall follow the traditional WKB method for tunneling.

    Now in order to solve the above equation, we make the following Hamilton--Jacobi ansatz
\begin{eqnarray}
A_{\m}=a_{\m}e^{-\frac{i}{\hbar}I(t,r,\theta,\phi)},\label{an1}
\end{eqnarray}
where $a_\mu$ is the polarization vector and $I$ is the action. With this ansatz, first and second order derivatives of $A_{\m}$  in (\ref{eom3}) can be written as,
\begin{eqnarray}
&&\p_{\s}A_{\m}=\Big(\p_{\s}a_{\m}-\frac{i}{\hbar}a_{\m}\p_{\s}I\Big)e^{-\frac{i}{\hbar}I}\label{an2}
\\
&&\p_{\r}\p_{\s}A_{\m}=\Big(\p_{\r}\p_{\s}a_{\m}-\frac{i}{\hbar}\p_{\r}a_{\m}\p_{\s}I-\frac{1}{\hbar^2}a_{\m}\p_{\r}I\p_{\s}I-\frac{i}{\hbar}a_{\m}\p_{\r}\p_{\s}I-\frac{i}{\hbar}(\partial_{\sigma} a_{\mu})(\partial_{\rho} I)\Big)e^{-\frac{i}{\hbar}I}.
\label{an3}
\end{eqnarray}
Substituting (\ref{an1}), (\ref{an2}) and (\ref{an3}) in (\ref{eom3}), we get the following equation
\begin{eqnarray}
&&g^{\rho\sigma}\left[(\hbar^2\p_{\rho}\p_{\sigma}a_{\m}-i\hbar\p_{\r}a_{\m}\p_{\s}I-a_{\m}\p_{\r}I\p_{\s}I-i\hbar\p_{\r}\p_{\s}I)\right]\nonumber\\&&-g^{\rho\sigma}\left[2\Gamma^{\lambda}_{\sigma\m}(\hbar^2\p_{\rho}a_{\lambda}-i\hbar a_{\lambda}\p_{\rho}I)+\Gamma^{\lambda}_{\r\s}(\hbar^2\p_{\lambda}a_{\m}-i\hbar a_{\m}\p_{\lambda}I)\right]\nonumber\\&&+\hbar^2g^{\rho\sigma}\left[-\p_{\r}\Gamma^{\lambda}_{\sigma\m}a_{\lambda}+(\Gamma^{\lambda}_{\r\s}\Gamma^{\a}_{\lambda\m}+\Gamma^{\lambda}_{\r\m}\Gamma^{\a}_{\s\lambda})a_{\a}\right]+\hbar^2R_{\m}^{\rho}a_{\rho}=0.\label{99}
\end{eqnarray}
Now we expand $I$ and $a_\mu$ in power series of $\hbar$
\begin{eqnarray}
&&I(r,t,\theta,\phi)=I_0(r,t,\theta,\phi)+\displaystyle\sum_{i=1}^{\infty} \hbar^i I_i(r,t,\theta,\phi)
\label{1.122}
\\
&&a_\mu=a_{\mu0}+\displaystyle\sum_{i=1}^{\infty} \hbar^i a_{\mu i}.
\label{1.13}
\end{eqnarray}
In the above expansions the terms $I_0$ and $a_{\mu0}$ are semiclassical values whereas the remaining terms are quantum corrections involving different powers of $\hbar$.
We substitute the above equation in (\ref{99}) and take the semiclassical limit  ($\hbar\rightarrow 0$)  to obtain
\begin{eqnarray}
g^{\r\s}a_{\mu0}(\p_{\r}I_0)(\p_{\s}I_0)=0.\label{rad}
\end{eqnarray}
Since the tunneling occurs in the radial direction, the ($r-t$) sector of the metric is relevant and in that case we write (\ref{rad}) as
\begin{eqnarray}
g^{tt}(\p_tI_0)^2+g^{rr}(\p_rI_0)^2=0.
\end{eqnarray}
For our choice of metric (\ref{ds2}), the above equation reduces to
\begin{eqnarray}
-\frac{1}{f}(\p_tI_0)^2+g(\p_rI_0)^2=0
\end{eqnarray}
which is equivalently written as
\begin{eqnarray}
\p_tI_0=\pm\sqrt{fg}\p_rI_0.\label{same}
\end{eqnarray}
This is the semiclassical Hamilton-Jacobi equation.
Now in order to find the Hamilton--Jacobi solution of $I_0$, we note that the metric (\ref{ds2}) that we have taken is stationary and so it has timelike Killing vectors. Thus we take the solution of (\ref{same}) in the form
\begin{eqnarray}
I_0(r,t,\theta,\phi)=\Omega t+\tilde I_0(r)+I_0'(\theta,\phi)\label{a1}
\end{eqnarray}
where $\Omega$ is the constant of motion corresponding to the timelike Killing vectors. In a general spacetime $\Omega$ is the product of the particle's energy $\omega$ as measured by an arbitrary observer and the appropriate redshift factor $V=\sqrt{-g_{tt}}$. Substituting this in (\ref{same}) we get,
\begin{eqnarray}
\Omega=\pm\sqrt{fg}\frac{d\tilde I_0}{dr}.
\end{eqnarray}
Integrating the above equation we find
\begin{eqnarray}
\tilde I_0(r)=\pm\Omega\int_0^r\frac{dr}{\sqrt{fg}}\label{a2}
\end{eqnarray}
where the limits of the integration are taken such that the particle passes through the horizon $r=r_H$. The $+(-)$ sign indicates that the particle is ingoing (outgoing). Combination of (\ref{a1}) and (\ref{a2}) gives the solution for $I_0(r,t)$
\begin{eqnarray}
I_0(r,t,\theta,\phi)=\Omega t\pm\Omega\int_0^r\frac{dr}{\sqrt{fg}}+I_0'(\theta,\phi).
\label{45}
\end{eqnarray}
Making use of the relations (\ref{an1}) and (\ref{45}) in the semiclassical limit we obtain the ingoing and outgoing solutions of the Maxwell equation in curved spacetime  
\begin{eqnarray}
A_{\mu{\textrm {(in)}}}\sim a_{\mu0}{\textrm{exp}}\Big[-\frac{i}{\hbar}\Big(\Omega t  +\Omega\int_0^r\frac{dr}{\sqrt{f(r)g(r)}}+I_0'(\theta,\phi)\Big)\Big]
\label{1.23}
\end{eqnarray} 
and
\begin{eqnarray}
A_{\mu{\textrm {(out)}}}\sim a_{\mu 0}{\textrm{exp}}\Big[-\frac{i}{\hbar}\Big(\Omega t  -\Omega\int_0^r\frac{dr}{\sqrt{f(r)g(r)}}+I_0'(\theta,\phi)\Big)\Big].
\label{1.24}
\end{eqnarray}
When a particle tunnels through the horizon, the sign of the metric coefficient in the $(r-t)$ sector changes. This suggests that there is an imaginary part in the time coordinate for the crossing of the black hole horizon and therefore a temporal contribution will appear in the expressions of probabilities for the ingoing and outgoing particles.

Thus the ingoing and outgoing probabilities of the particle are given by,
\begin{eqnarray}
P_{{\textrm{in}}}=|A_{\mu{\textrm {(in)}}}|^2\sim {\textrm{exp}}\Big[\frac{2}{\hbar}\Big(\Omega{\textrm{Im}}~t +\Omega{\textrm{Im}}\int_0^r\frac{dr}{\sqrt{f(r)g(r)}}\Big)\Big]
\label{1.25}
\end{eqnarray}
and
\begin{eqnarray}
P_{{\textrm{out}}}=|A_{\mu{\textrm {(out)}}}|^2\sim {\textrm{exp}}\Big[\frac{2}{\hbar}\Big(\Omega{\textrm{Im}}~t -\Omega{\textrm{Im}}\int_0^r\frac{dr}{\sqrt{f(r)g(r)}}\Big)\Big].
\label{1.26}
\end{eqnarray}
Note that the angular contribution $I_0'(\theta,\phi)$ does not appear in the above expressions of probabilities.
In the limit $\hbar\rightarrow 0$, everything is absorbed in the black hole and hence the ingoing probability $P_{\textrm{in}}$ must be unity. Therefore, in this limit, (\ref{1.25}) yields,
\begin{eqnarray}
{\textrm{Im}}~t = -{\textrm{Im}}\int_0^r\frac{dr}{\sqrt{f(r)g(r)}}.
\label{1.27}
\end{eqnarray}
It must be noted that the above relation satisfies the classical condition $\frac{\partial I_0}{\partial\Omega}=$ constant. This is understood by the following argument. Calculating the left side of this condition from (\ref{45}) we obtain,
\begin{eqnarray}
t = \mp \int_0^r\frac{dr}{\sqrt{f(r)g(r)}}
\label{time1}
\end{eqnarray}
where $-(+)$ sign indicates that the particle is ingoing (outgoing). So for an ingoing particle this condition immediately yields (\ref{1.27}). On the other hand a naive substitution of `Im$~t$' in (\ref{1.26}) from (\ref{time1}) for the outgoing particle gives $P_{{\textrm{out}}}=1$. But it must be noted that according to classical general theory of relativity, a particle can be absorbed in the black hole, while the reverse process is forbidden. In this regard, ingoing classical trajectory exists while the outgoing classical trajectory is forbidden. Hence use of the classical condition for outgoing particle is meaningless.

    Now to find out `Im$~t$' for the outgoing particle, we will take the help of the Kruskal coordinates which are well behaved throughout the space-time.
The Kruskal time ($T$) and space ($X$) coordinates
inside and outside the horizon are defined as \cite{Ray}
\begin{eqnarray}
&&T_{is}=e^{\kappa r^{*}_{is}} \cosh\!\left(\kappa t_{is}\right)~~;\hspace{4ex}
X_{is} = e^{\kappa r^{*}_{is}} \sinh\!\left(\kappa t_{is}\right)
\label{Krus1.1}
\\
&&T_{os}=e^{\kappa r^{*}_{os}} \sinh\!\left( \kappa t_{os}\right)~~;\hspace{4ex}
X_{os} = e^{\kappa r^{*}_{os}}  \cosh\!\left(\kappa t_{os}\right)
\label{Krus1.2}
\end{eqnarray}
where $\kappa$ is the surface gravity defined by
\begin{eqnarray}
\kappa = \frac{1}{2} \sqrt{f'(r_H)g'(r_H)}~.
\end{eqnarray}
Here `$is(os)$' stands for the inside (outside) the event horizon while $r^*$ is the tortoise coordinate, defined by
\begin{eqnarray}
r^* = \int\frac{dr}{\sqrt{f(r)g(r)}} ~.
\label{tor}
\end{eqnarray}
These two sets of coordinates are connected through the following relations
\begin{eqnarray}
&& t_{is} = t_{os}-i\frac{\pi}{2\kappa}
\label{Krus2.1}\\
&& r^{*}_{is} = r^{*}_{os} + i\frac{\pi}{2\kappa}
\label{Krus2.2}
\end{eqnarray}
so that the Kruskal coordinates get identified as $T_{is} = T_{os}$ and $X_{is} = X_{os}$. This indicates that when a particle travels from inside to outside the horizon, `$t$' coordinate picks up an imaginary term $-\frac{\pi}{2{\kappa}}$. This is precisely given by (\ref{1.27}). It is noteworthy that exactly the same imaginary temporal contribution was needed to solve the problem of factor $2$ in the expression of black hole temperature. This was first proposed in \cite{dog}. A more elaborate discussion on the method we follow in the present paper may be found in \cite{RB,vagenas}.

  Therefore, using (\ref{1.27}) in (\ref{1.26}) we get the probability for the outgoing particle 
\begin{eqnarray}
P_{{\textrm{out}}}\sim{\textrm{exp}}\Big[-\frac{4}{\hbar}\Omega{\textrm{Im}}\int_0^r\frac{dr}{\sqrt{f(r)g(r)}}\Big].
\label{1.28}
\end{eqnarray}
Now if an observer at infinity (i.e. $r\rightarrow\infty$) observes the same tunneling process (corresponding to Hawking effect) with energy $\omega$ and temperature $T_H$, then it reads the principle of ``detailed balance'' as
\begin{eqnarray}
\frac{P_{{\textrm{out}}}}{P_{{\textrm{in}}}}=e^{-\omega/T_H}.
\end{eqnarray}
Since $P_{{\textrm{in}}}=1$, the above equation leads to
\begin{eqnarray}
P_{{\textrm{out}}}=e^{-\omega/T_H}.
\label{1.29}
\end{eqnarray}
Now at $r\rightarrow\infty$, $\sqrt{-g_{tt}}=1$ and so $\Omega=\omega$. Therefore comparing (\ref{1.28}) and (\ref{1.29}) we get the black hole temperature as
\begin{eqnarray}
T_H =\frac{\hbar}{4}\Big[{\textrm{Im}}\int_0^r\frac{dr}{\sqrt{f(r)g(r)}}\Big]^{-1}.
\label{1.30}
\end{eqnarray}
This is the standard Hawking temperature obtained earlier by using tunneling method of scalar \cite{Majhibeyond} or Dirac \cite{Majhifermi} particle. This confirms that a black hole can radiate any type of particle like a black body.

\subsection{Method 2}
In this subsection we study the same problem, namely, the tunneling of photon using Hamilton--Jacobi method, but taking the action (\ref{77}) without the gauge fixing term. So our action reads
\begin{eqnarray}
S=-\frac{1}{4}\int F_{\m\n} F^{\m\n}\sqrt{-g}d^4x\label{S}
\end{eqnarray}
and we take care of the gauge invariance of the theory by imposing the Lorentz gauge condition later. An arbitrary variation of $A_{\m}$ in the action (\ref{S}) gives the equation of motion
\begin{eqnarray}
\nabla^{\m}F_{\m\n}=0.
\end{eqnarray}
Using the standard method of calculating the covariant derivative of a tensor field we write the above equation as,
\begin{eqnarray}
g^{\n\a}[\p_{\a}F_{\m\n}-\Gamma^{\lambda}_{\a\m}F_{\lambda\nu}-\Gamma^{\lambda}_{\a\n}F_{\mu\lambda}]=0.\label{max}
\end{eqnarray}

 Using the definition of the field tensor $F_{\mu\nu}$ (\ref{fmunu}) in the above equation and then substituting expressions (\ref{an2}) and (\ref{an3}) together with expansions (\ref{1.122}) and (\ref{1.13}) we get order $\mathcal O(\hbar^0)$ equation as
\begin{eqnarray}
g^{\nu\alpha}(-a_{\nu 0}\partial_{\alpha}I_0\partial_{\mu}I_0+a_{\mu 0}\partial_{\alpha}I_0\partial_{\nu}I_0)=0.
\label{max1}
\end{eqnarray}
This is not the semiclassical Hamilton-Jacobi equation (\ref{same}). Also it is not possible to obtain solutions for $I_0(r,t)$ in terms of metric coefficients. Therefore,
in order to proceed further, now we impose the Lorentz gauge in the curved spacetime
\begin{eqnarray}
\p_{\m}\big(\sqrt{-g}A^{\m}\big)=0.
\label{gauge1}
\end{eqnarray}
This can be equivalently written as,
\begin{eqnarray}
\nabla^{\m}A_{\m}\equiv
g^{\m\n}(\p_{\n}A_{\m}-\Gamma^{\s}_{\n\m}A_{\s})=0.\label{gaugecondition}
\end{eqnarray}
Again using (\ref{an2},\ref{an3}) and (\ref{1.122},\ref{1.13}) in the above equation and comparing $\hbar^0$ order terms on both sides we find, 
\begin{eqnarray}
g^{\nu\alpha}a_{\nu 0}\partial_{\alpha}I_0=0.
\label{con0}
\end{eqnarray}
Due to (\ref{con0}), (\ref{max1}) simplifies to (\ref{rad}) which ultimately gives the desired semiclassical Hamilton-Jacobi equation (\ref{same}) obtained in the previous subsection. Rest of the analysis to find the Hawking temperature is identical to the previous study. Naturally, the resulting black hole temperature is found to be (\ref{1.30}).

\section{Correction to the semiclassical results}
So far our analysis was restricted only upto semiclassical approximation. In the present section we shall study the effects of quantum corrections on the black hole temperature for the tunneling of photon. To do this, we can follow either of the methods discussed in the previous section. Since the calculation based on first method is found to be more complicated, here we follow the second method and improve the previous analysis by incorporating the first order quantum effects.

Substituting (\ref{1.122}) and (\ref{1.13}) in (\ref{max}) and then equating first order quantum correction ($\mathcal O(\hbar^1)$) on both sides, we find
\begin{eqnarray}
&&g^{\nu\alpha}\Big[-i\partial_{\alpha}a_{\nu 0}\partial_{\mu}I_0-a_{\nu 1}\partial_{\alpha}I_0\partial_{\mu}I_0-a_{\nu 0}\partial_{\alpha}I_1\partial_{\mu}I_0-a_{\nu 0}\partial_{\alpha}I_0\partial_{\mu}I_1
\nonumber
\\
&+&a_{\mu 1}\partial_{\alpha}I_0\partial_{\nu}I_0+a_{\mu 0}\partial_{\alpha}I_1\partial_{\nu}I_0+a_{\mu 0}\partial_{\alpha}I_0\partial_{\nu}I_1+i\partial_{\alpha}a_{\mu 0}\partial_{\nu}I_0 \Big]\nonumber\\
&-&g^{\nu\alpha}\Gamma^{\lambda}_{\alpha\mu}[-ia_{\nu 0}\partial_{\lambda}I_0+ia_{\lambda 0}\partial_{\nu}I_0]-g^{\nu\alpha}\Gamma^{\lambda}_{\alpha\nu}[-ia_{\lambda 0}\partial_{\mu}I_0+ia_{\mu 0}\partial_{\lambda}I_0]=0.
\label{hbar1}
\end{eqnarray}
Using (\ref{con0}) and (\ref{rad}) we simplify (\ref{hbar1}) to get
\begin{eqnarray}
&&-ig^{\nu\alpha}(\partial_{\mu}I_0)\Big[\partial_{\alpha}a_{\nu 0}-ia_{\nu 1}\partial_{\alpha}I_0-ia_{\nu 0}\partial_{\alpha}I_1-\Gamma^{\lambda}_{\alpha\nu}a_{\lambda 0}\Big]
\nonumber
\\
&+&g^{\nu\alpha}\Big[i\partial_{\alpha}a_{\mu 0}\partial_{\nu}I_0+a_{\mu 0}\partial_{\alpha}I_1\partial_{\nu}I_0+a_{\mu 0}\partial_{\alpha}I_0\partial_{\nu}I_1 \Big]\nonumber\\
&-&g^{\nu\alpha}\Gamma^{\lambda}_{\alpha\mu}[-ia_{\nu 0}\partial_{\lambda}I_0+ia_{\lambda 0}\partial_{\nu}I_0]-ig^{\nu\alpha}\Gamma^{\lambda}_{\alpha\nu}a_{\mu 0}\partial_{\lambda}I_0]=0.
\label{main2}
\end{eqnarray}
This equation alone is not sufficient to find the solution of $I_1$. As before we need to use the gauge condition (\ref{gaugecondition}). Substituting (\ref{1.122}) and (\ref{1.13}) in (\ref{gaugecondition}) and then equating the terms of the order of $\hbar^1$ on both sides we get
\begin{eqnarray}
g^{\nu\alpha}\Big(\partial_{\alpha}a_{\nu 0}-ia_{\nu 1}\partial_{\alpha}I_0-ia_{\nu 0}\partial_{\alpha}I_1-\Gamma^{\lambda}_{\alpha\nu}a_{\lambda 0}  \Big)=0.
\label{con1}
\end{eqnarray}
Using (\ref{con1}) in (\ref{main2}) we obtain
\begin{eqnarray}
&&g^{\nu\alpha}\Big[i\partial_{\alpha}a_{\mu 0}\partial_{\nu}I_0+a_{\mu 0}\partial_{\alpha}I_1\partial_{\nu}I_0+a_{\mu 0}\partial_{\alpha}I_0\partial_{\nu}I_1 \Big]\nonumber\\
&-&g^{\nu\alpha}\Gamma^{\lambda}_{\alpha\mu}[-ia_{\nu 0}\partial_{\lambda}I_0+ia_{\lambda 0}\partial_{\nu}I_0]-ig^{\nu\alpha}\Gamma^{\lambda}_{\alpha\nu}a_{\mu 0}\partial_{\lambda}I_0]=0.
\label{main4}
\end{eqnarray}
Since only ($r-t$) sector of the metric is relevant in our analysis, the above expression can be expanded as
\begin{eqnarray}
&&-\frac{1}{f}\Big[i\partial_{t}a_{\mu 0}\partial_{t}I_0+2a_{\mu 0}\partial_{t}I_1\partial_{t}I_0\Big]+g\Big[i\partial_{r}a_{\mu 0}\partial_{r}I_0+2a_{\mu 0}\partial_{r}I_1\partial_{r}I_0\Big]\nonumber\\
&&-i\frac{1}{f}\Gamma^{r}_{t\mu}[a_{t 0}\partial_{r}I_0-a_{r 0}\partial_{t}I_0]+ig\Gamma^{t}_{r\mu}[a_{r 0}\partial_{t}I_0-a_{t 0}\partial_{r}I_0]\nonumber\\
&&+\frac{i}{f}\Gamma^{r}_{tt}a_{\mu 0}\partial_{r}I_0-ig\Gamma^r_{rr}a_{\mu 0}\partial_rI_0=0.
\end{eqnarray}
Making use of (\ref{same}) and ($r-t$) component of (\ref{con0}) we reduce the above equation as
\begin{eqnarray}
&&-\frac{1}{f}[\pm i\sqrt{fg}\partial_ta_{\mu 0}\pm 2a_{\mu 0}\sqrt{fg}\partial_tI_1]+g[i\partial_ra_{\mu 0}+2a_{\mu 0}\partial_rI_1]\nonumber\\
&&+\frac{i}{f}\Gamma^{r}_{tt}a_{\mu 0}-ig\Gamma^r_{rr}a_{\mu 0}=0.
\label{final}
\end{eqnarray}
Since the terms independent of the single particle action $I$ will not contribute to the thermodynamic quantities, we drop them from (\ref{final}) to find,
\begin{eqnarray}
\p_tI_1=\pm \sqrt{fg}\p_rI_1.
\label{eqn}
\end{eqnarray}
This equation is quite analogous to its semiclassical counterpart (\ref{same}). 
Comparing this with (\ref{same}) we get
\begin{eqnarray}
(\p_tI_i)=\pm\sqrt{fg}(\p_r I_i)\label{ii}
\end{eqnarray}
for $i=0$ and 1. This implies that the solution of these equations are not independent and $I_1$ is proportional to $I_0$. Thus (\ref{1.122}) is written as
\begin{eqnarray}
I(r,t,\theta,\phi)=I_0(r,t,\theta,\phi)+\hbar\gamma I_0(r,t,\theta,\phi)\label{expansion}
\end{eqnarray}
where $\gamma$ is the proportionality constant. Since the action $I$ has the dimension of $\hbar$, the proportionality constant ($\gamma$) should have the dimension of $\hbar^{-1}$. Again in our units $G=c=k_B=1$ and $\hbar$ has the dimension of mass square. So $\gamma$ is of the form
\begin{eqnarray}
\gamma=\frac{\beta_1}{M^{2}},
\label{gamma}
\end{eqnarray}
where $M$ is the mass of the black hole, the only mass parameter that appears in the problem. $\beta_1$ is some dimensionless constant having value such that quantum correction is of the order of $\hbar$. Combining (\ref{expansion}) and (\ref{gamma}) we get
\begin{eqnarray}
I=\big(1+\beta_1\frac{\hbar}{M^{2}}\big)I_0.\label{ii0}
\end{eqnarray}
Therefore to obtain a solution of $I$ upto $\hbar^1$ order, it is sufficient to solve $I_0$.
The solution for $I_0$ was obtained in the previous section which is given by (\ref{45}). Substituting (\ref{45}) in (\ref{ii0}) we get the action
\begin{eqnarray}
I=\big(1+\beta_1\frac{\hbar}{M^{2}}\big)\left[\Omega t\pm\Omega\int_0^r\frac{dr}{\sqrt{fg}}+I_0'(\theta,\phi)\right].\label{nw}
\end{eqnarray}
Above equation contains the quantum correction together with the standard semiclassical term. Expectedly in the limit $\hbar\rightarrow 0$, (\ref{nw}) reduces to (\ref{45}).

Having obtained the solution of the single particle action, we can follow the analysis of subsection 2.1 in a straight forward manner to calculate the black hole temperature. The modified Hawking temperature upto first order quantum correction thus obtained is
\begin{eqnarray}
T_h&=&\frac{\hbar}{4}\Big[\big(1+\beta_1\frac{\hbar}{M^{2}}\big){\textrm{Im}}\int_0^r\frac{dr}{\sqrt{f(r)g(r)}}\Big]^{-1}
\nonumber
\\
&=&T_H\Big(1+\beta_1\frac{\hbar}{M^{2}}\Big)^{-1}
\label{1.50}
\end{eqnarray}
where $T_H$ is the semiclassical Hawking temperature given by (\ref{1.30}). This expression exactly matches with earlier results upto $\hbar^1$ order for scalar \cite{Majhibeyond} or Dirac \cite{Majhifermi} particle tunneling.

\section{Gravitino tunneling beyond semiclassical approximation}
We follow the method discussed in the previous section to study gravitino tunneling. As claimed in the introduction, our analysis goes beyond the semiclassical approximation by incorporating all possible quantum corrections. The semiclassical Hawking temperature is shown to be altered properly.

The Rarita-Schwinger equation\cite{Ra} for the massless spin-$3/2$ fermion in a curved spacetime background is given by
\begin{eqnarray}
i\gamma^\mu\nabla_\mu\psi_{\nu}=0,
\label{1.02}
\end{eqnarray}
together with a constraint
\begin{eqnarray}
\gamma^{\m}\psi_{\m}=0\label{mixd}
\end{eqnarray}
to ensure that there is no Dirac state in $\psi$. Here $\psi_{\nu}\equiv\psi_{\nu a}$ is a vector valued spinor and the covariant derivative is defined in the usual way,
\begin{eqnarray}
&&\nabla_\mu=\partial_\mu+\frac{i}{2}\Gamma{^\alpha}{_\mu}{^\beta}\Sigma_{\alpha\beta};\,\,\ \Gamma{^\alpha}{_\mu}{^\beta}=g^{\beta\nu}\Gamma^\alpha_{\mu\nu};\,\,\ \Sigma_{\alpha\beta}=\frac{i}{4}\Big[\gamma_\alpha,\gamma_\beta\Big].
\label{1.04}
\end{eqnarray} 
We take the following representations of the $\gamma$ matrices 
\begin{eqnarray}
\gamma^t&=&\frac{1}{\sqrt{f(r)}}\left(\begin{array}{c c}
i & 0 \\
0 & -i
\end{array}\right);\,\,\
\gamma^r=\sqrt{g(r)}\left(\begin{array}{c c}
0 & \sigma^3 \\
\sigma^3 & 0
\end{array}\right)
\nonumber
\\
\gamma^\theta&=&\frac{1}{r}\left(\begin{array}{c c}
0 & \sigma^1 \\
\sigma^1 & 0
\end{array}\right);\,\,\,\
\gamma^\phi=\frac{1}{r\textrm{sin}\theta}\left(\begin{array}{c c}
0 & \sigma^2 \\
\sigma^2 & 0
\end{array}\right)
\label{1.03}
\end{eqnarray}
which satisfy $\{\gamma^\mu,\gamma^\nu\}=2g^{\mu\nu}$. Since we are working only with the radial trajectories, the $(r-t)$ sector of the metric (\ref{ds2}) is important. Hence (\ref{1.02}) is expressed as
\begin{eqnarray}
i\gamma^\mu\partial_\mu\psi_{\nu}-\frac{1}{2}\Big(g^{tt}\gamma^\mu\Gamma^r_{\mu t}-g^{rr}\gamma^\mu\Gamma^t_{\mu r}\Big)\Sigma_{rt}\psi_{\nu}=0.
\label{1.05}
\end{eqnarray}
Substituting the metric coefficients and the non-vanishing connections
\begin{eqnarray}
\Gamma^r_{tt}=\frac{f'g}{2};\,\,\ \Gamma^t_{tr}=\frac{f'}{2f}
\label{1.06}
\end{eqnarray}
for the metric (\ref{ds2}) in (\ref{1.05}), we get the following equation
\begin{eqnarray}
i\gamma^t\partial_t\psi_{\m}+i\gamma^r\partial_r\psi_{\m}+\frac{f'g}{2f}\gamma^t\Sigma_{rt}\psi_{\m}=0.
\label{1.07}
\end{eqnarray}
We take the following ansatz for the wave function
\begin{eqnarray}
\psi_{\mu}(t,r)= \left(\begin{array}{c}
A_{\mu}(t,r) \\
B_{\mu}(t,r) \\
C_{\mu}(t,r) \\
D_{\mu}(t,r)
\end{array}\right)=\left(\begin{array}{c}
a_{\mu} \\
b_{\mu} \\
c_{\mu} \\
d_{\mu}
\end{array}\right){\textrm{exp}}\Big[-\frac{i}{\hbar}I(t,r)\Big]
\end{eqnarray}
where $I(r,t)$ is the action. Using this ansatz and calculating the
value of $\Sigma$ from (\ref{1.04})
\begin{eqnarray}
\Sigma_{rt}=\frac{i}{2}\left(\begin{array}{c c c c}
0 & 0 & i\sqrt{\frac{f}{g}} & 0\\ 
0 & 0 & 0 & -i\sqrt{\frac{f}{g}}\\
-i\sqrt{\frac{f}{g}} & 0 & 0 & 0\\
0 & i\sqrt{\frac{f}{g}} & 0 & 0
\end{array}\right),
\label{1.08}
\end{eqnarray}
we write (\ref{1.07}) component-wise as,
\begin{eqnarray}
\frac{\hbar}{\sqrt{f}}(\p_ta_{\m})+\frac{i}{\sqrt{f}}a_{\m}(\p_tI)-i\hbar\sqrt{g}(\p_rc_{\m})+\sqrt{g}c_{\m}(\p_rI)-\frac{\hbar f'\sqrt{g}}{2f}c_{\m}=0\label{gravitino1}\\
\frac{\hbar}{\sqrt{f}}(\p_tb_{\m})+\frac{i}{\sqrt{f}}b_{\m}(\p_tI)+i\hbar\sqrt{g}(\p_rd_{\m})-\sqrt{g}d_{\m}(\p_rI)+\frac{\hbar f'\sqrt{g}}{2f}d_{\m}=0\\
-\frac{\hbar}{\sqrt{f}}(\p_tc_{\m})-\frac{i}{\sqrt{f}}c_{\m}(\p_tI)-i\hbar\sqrt{g}(\p_ra_{\m})+\sqrt{g}a_{\m}(\p_rI)-\frac{\hbar f'\sqrt{g}}{2f}a_{\m}=0\\
-\frac{\hbar}{\sqrt{f}}(\p_td_{\m})-\frac{i}{\sqrt{f}}d_{\m}(\p_tI)+i\hbar\sqrt{g}(\p_rb_{\m})-\sqrt{g}b_{\m}(\p_rI)+\frac{\hbar f'\sqrt{g}}{2f}b_{\m}=0.
\label{gravitino2}
\end{eqnarray}
Here, we ignore the constraint equation (\ref{mixd}) since they are not important for the solution of the action. In the above equations, the terms which do not involve the single particle action will not contribute to the thermodynamic entities of the black hole. Therefore we drop those terms to write (\ref{gravitino1})--(\ref{gravitino2}) as
\begin{eqnarray}
-\frac{i}{\sqrt{f}}a_{\m}(\p_tI)-\sqrt{g}c_{\m}(\p_rI)=0\label{grav1}\\
-\frac{i}{\sqrt{f}}b_{\m}(\p_tI)+\sqrt{g}d_{\m}(\p_rI)=0\label{grav2}\\
\frac{i}{\sqrt{f}}c_{\m}(\p_tI)-\sqrt{g}a_{\m}(\p_rI)=0\label{grav3}\\
\frac{i}{\sqrt{f}}d_{\m}(\p_tI)+\sqrt{g}b_{\m}(\p_rI)=0.
\label{grav4}\end{eqnarray}
From (\ref{grav1}) and (\ref{grav3}) we note that $a_{\m}$ and $c_{\m}$ will have nonvanishing values only when
\begin{eqnarray}
{\textrm{det}}\left(\begin{array}{c c}
-\frac{i}{\sqrt{f}}(\p_tI) & -\sqrt{g}(\p_rI)\\ 
-\sqrt{g}(\p_rI) & \frac{i}{\sqrt{f}}(\p_tI)\end{array}\right)=0.
\end{eqnarray}
 This condition gives the result
\begin{eqnarray}
(\p_tI)^2=fg(\p_rI)^2
\end{eqnarray}
or equivalently,
\begin{eqnarray}
\p_tI=\pm\sqrt{fg}\p_rI.\label{pm}
\end{eqnarray}
Substituting (\ref{pm}) in (\ref{grav1}) we get
\begin{eqnarray}
a_{\m}=\pm ic_{\m}.\label{pm1}
\end{eqnarray}
The above results can also be obtained by solving (\ref{grav2}) and (\ref{grav4}) simultaneously. 
As before, we expand $I,a_{\m},$ and $c_{\m}$ in power series of $\hbar$:
\begin{eqnarray}
&&I(r,t)=I_0(r,t)+\displaystyle\sum_{i=1}^{\infty} \hbar^i I_i(r,t)
\label{expansion1}
\\
&&a_{\m}=a_{\m 0}+\displaystyle\sum_{i=1}^{\infty} \hbar^i a_{\m i}; \,\,\, c_{\m}=c_{\m 0}+\displaystyle\sum_{i=1}^{\infty} \hbar^i c_{\m i}.
\label{expansion2}
\end{eqnarray}
Now substituting these in (\ref{pm}) and (\ref{pm1}) and then equating different powers of $\hbar$ on both sides of equation we obtain,
\begin{eqnarray}
\p_tI_i=\pm\sqrt{fg}\p_rI_i
\label{pm2}
\end{eqnarray}
and
\begin{eqnarray}
a_{\m i}=\pm ic_{\m i}\label{pm3}
\end{eqnarray}
for $i=0,1,2,\cdot\cdot\cdot$.
Note that (\ref{pm2}) is same as (\ref{ii}) for $i=0,1$ which was obtained order by order for the photon field.

Now following the analysis presented in the earlier sections, we can calculate the black hole temperature due to gravitino tunneling. The result thus obtained is\begin{eqnarray}
T_h=T_H\big(1+\sum_{i=1}^{\infty}\beta_{i}\frac{\hbar^i}{M^{2i}}\big)^{-1}
\end{eqnarray}
which upon first order approximation matches with (\ref{1.50}), though the values of the first order correction parameter $\beta_1$ for photon and gravitino are not same. This point will be examined in detail in the next section.

    Some comments on the corrected form of the Hawking temperature (\ref{1.50}) are as follows. For the Schwarzschild black hole $f(r)=g(r)=1-\frac{2M}{r}$. Substituting this in (\ref{1.50}) and performing the contour integration we obtain the first order quantum corrected Hawking temperature as
\begin{eqnarray}
T_h=\frac{\hbar}{8\pi M}\Big(1+\beta_1\frac{\hbar}{M^2}\Big)^{-1}.
\label{b}
\end{eqnarray}
Using this corrected form of temperature and exploiting the Hawking's periodicity arguments one can find the corrected form of the Schwarzschild metric upto $\hbar$ order as
\begin{eqnarray}
ds^2_{{\textrm{corr}}}=-\Big[1-\frac{2M}{r}\Big(1+\beta_1\frac{\hbar}{M^2}\Big)\Big]dt^2+\frac{dr^2}{\Big[1-\frac{2M}{r}\Big(1+\beta_1\frac{\hbar}{M^2}\Big)\Big]}+r^2d\Omega^2.
\label{aa1}
\end{eqnarray}
For detailed discussions see \cite{Majhiback}. Therefore the fractional change of the metric coefficients is $-\frac{\beta_1\hbar}{M^2}$ which is precisely the result given in \cite{JM}. The previous derivation was based on the solution of Einstein equation including the renormalized energy-momentum tensor for the back reaction effect in the spacetime. In that case the coefficient (which is $\beta_1$ for our case) is related to the number of different types of fields. In the next section we shall explicitly show how our result matches with earlier work \cite{York,Lousto}  which incorporates the effect of all loops back reaction in the spacetime.

      Now from the first law of thermodynamics $dS_{\textrm{bh}}=\frac{dM}{T_h}$, it is easy to find the corrected form of the Bekenstein-Hawking entropy which in this case is given by,
\begin{eqnarray}
S_{\textrm{bh}}=\frac{A}{4\hbar}+4\pi\beta_1\ln A+{\textrm{higher order terms in $\hbar$}}
\label{aa2}
\end{eqnarray}
where $A=16\pi M^2$ is the area of the event horizon of the Schwarzschild black hole. The first term is the usual semiclassical result and the second term is the logarithmic correction \cite{Fursaev,Page,Kaul,veg,Majhi,Majhibeyond} which in this case comes from $\hbar$ order correction to the one particle action and so on. In the next section we will discuss a method of fixing the coefficients.

\section{Discussions on correction parameter $\beta_1$}
         In this section we discuss about the undetermined coefficient $\beta_1$ for both photon and gravitino cases. To determine this, we begin by studying the behaviour of actions (\ref{ii0}) and (\ref{expansion1}) for the photon tunneling first. Apparently, one might think, for a zero rest mass field the trace of the energy-momentum tensor ($T^\mu_\mu$) is zero. But the point is, at the quantum level it is not possible to preserve the conformal and diffeomorphism symmetries simultaneously. In fact, violation of the conformal invariance leads to a nonvanishing $T^\mu_\mu$. For chiral theory, both of these symmetries are violated and therefore both divergence and trace of energy-momentum tensor are nonzero. This point has been rigorously studied for black hole case in \cite{Iso}. Throughout our analysis diffeomorphism symmetry is always preserved and so we connect $\beta_1$ only with the trace anomaly. This is done by simple scaling argument which was originally initiated by Hawking \cite{REF1}.

     Under an infinitesimal constant scale transformation, parametrized by $k$, 
the metric coefficients change as,
\begin{eqnarray}
\tilde{g}{_{\mu\nu}} = k g_{\mu\nu}\simeq(1+\delta k)g_{\mu\nu}.
\label{coff1}
\end{eqnarray}
Due to this transformation, the coefficients of $(r-t)$ sector of the metric (\ref{ds2}) change as $\tilde f=kf,\tilde g=k^{-1}g$. Also, to preserve the scale invariance of the Lorentz gauge condition (\ref{gauge1}), the field $A^\mu$ transforms as $\tilde {A^\mu}=k^{-2}A^\mu$. On the other hand, the action (\ref{S}) for photon field shows that $A^\mu$ has the dimension of mass. Since the only mass parameter we have in this problem is the black hole mass $M$, the infinitesimal change of it is given by,
\begin{eqnarray}
\tilde{M}=k^{-2}M\simeq(1-2\delta k)M.
\label{coff2}
\end{eqnarray}
Now from (\ref{45}) and (\ref{1.27}) the imaginary part of the semiclassical contribution of the outgoing single particle action is
\begin{eqnarray}
\textrm{Im}I{_0}{_{(\textrm{out})}} = -2\Omega{\textrm{Im}}~\int_0^r\frac{dr}{\sqrt{f(r)g(r)}}
\label{coff3}
\end{eqnarray}
where for $r\rightarrow \infty$, $\Omega=\omega$ which gets identified with the energy (i.e. mass $M$) of a stable black hole \cite{Majhi}. Therefore $\omega$  transforms according to $M$ under (\ref{coff1}).
Considering the imaginary part of the term $\mathcal O(\hbar)$ in (\ref{nw}), we get, under the scale transformation,      
\begin{eqnarray}
{\tilde{\cal{A}}}_{(1)}&\equiv&{\hbar}\textrm{Im}\tilde{I}{_1}{_{(\textrm{out})}}
=\Big(\frac{{\hbar}\beta_1}{\tilde{M}^2}\Big)\textrm{Im}\tilde I{_0}{_{(\textrm{out})}}.\nonumber
\end{eqnarray} 
Using (\ref{coff2}) we write the above equation as 
\begin{eqnarray}
{\tilde{\cal{A}}}_{(1)}
&\simeq&\Big(\frac{\hbar\beta_1}{{M}^2}\Big)(1+2\delta k)\textrm{Im} I{_0}{_{(\textrm{out})}}
\nonumber
\\
&=&{\cal{A}}_{(1)}+\Big(\frac{\hbar\beta_1}{{M}^2}\Big)2\delta k\textrm{Im} I{_0}{_{(\textrm{out})}}.
\label{coff4}
\end{eqnarray} 
Therefore the change of ${\cal{A}}_{(1)}$ is given by,
\begin{eqnarray}
\delta {\cal{A}}_{(1)}&=&{\tilde{\cal{A}}}_{(1)}-{\cal{A}}_{(1)}
\nonumber
\\
&\simeq&\Big(\frac{\hbar\beta_1}{{M}^2}\Big)2\delta k\textrm{Im} I{_0}{_{(\textrm{out})}}
\label{coff5}
\end{eqnarray}
which leads to the following equation,
\begin{eqnarray}
\frac{\delta {\cal{A}}_{(1)}}{\delta k}=2\Big(\frac{\hbar\beta_1}{{M}^2}\Big)\textrm{Im} I{_0}{_{(\textrm{out})}}.
\label{coff6}
\end{eqnarray}
At this point we use of the definition of energy-momentum tensor in the above equation to get,
\begin{eqnarray}
\textrm{Im}\int d^4x\sqrt{-g} T_\mu^\mu = \frac{2\delta {\cal{A}}_{(1)}}{\delta k}
=4\Big(\frac{\hbar\beta_1}{{M}^2}\Big)\textrm{Im} I{_0}{_{(\textrm{out})}}.
\label{coff7}
\end{eqnarray}
From (\ref{coff7}) it is clear that, in the presence of trace anomaly, the action is not invariant under the scale transformation. Since for the Schwarzschild black hole $f(r)=g(r)=1-\frac{2M}{r}$, from (\ref{coff3}) we obtain $\textrm{Im}I_{0{(\textrm{out})}}=-4\pi\omega M$. Substitution of this result in (\ref{coff7}) for $\omega=M$ we find
\begin{eqnarray}
\hbar\beta_1= -\frac{1}{16\pi}\textrm{Im}\int d^4x\sqrt{-g}T_\mu^\mu.
\label{coff8}
\end{eqnarray}  
Since the higher loop calculations to get $T_{\mu\nu}$ (from which $T_\mu^\mu$ is obtained) is very much complicated, usually in literature \cite{Witt} only one loop calculation for $T_{\mu\nu}$ is discussed. Thus, comparing only the $\hbar^1$ order on both sides of (\ref{coff8}), we obtain,
\begin{eqnarray}
\beta_1=-\frac{1}{16\pi}{\textrm{Im}}\int d^4x\sqrt{-g}{T^\mu_\mu}^{(1)}.
\label{coff9}
\end{eqnarray}
This relation clearly shows that $\beta_1$ is connected to the trace anomaly.

The correction to the black hole entropy (which is proportional to $\beta_1$) was calculated by Hawking himself and he showed it to be related to the trace anomaly \cite{REF1}. This was done by path integral approach based on zeta function regularization where the path integral was modified by taking into account the effect of fluctuations coming from the scalar field. The entropy expression found was
\begin{eqnarray}
S_{\textrm{bh}}=\frac{A}{4\hbar}-\frac{1}{2}\Big({\textrm{Im}}\int d^4x\sqrt{-g}T^\mu_\mu\Big)\ln A
\label{ref1}
\end{eqnarray}
which is equivalent to our result (\ref{aa2}). The coefficient of the logarithmic term of the above expression matches with (\ref{coff9}) apart from a numerical factor. This mismatch in the numerical factor is a consequence of the fact that we have considered the photon field instead of scalar field. Previously, it has been established \cite{Morette} that upto order $\hbar$, the result obtained from WKB ansatz are equivalent to the path integral result. Therefore, it is not surprising that the result obtained here from simple scaling arguments, is consistent with the path integral approach. 

    Following the identical analysis for the gravitino case one can immediately show that
\begin{eqnarray}
\beta_1|_{\textrm{gravitino}}=\frac{3}{8\pi}{\textrm{Im}}\int d^4x\sqrt{-g}{T_\mu}{^\mu}^{(1)}|_{\textrm{gravitino}}
\label{coff11}
\end{eqnarray}
where  ${T_\mu}{^\mu}^{(1)}|_{\textrm{gravitino}}$ is the trace of the renormalized energy-momentum tensor of gravitino upto one loop expansion.

    Similar relations were given earlier for scalar particle \cite{Majhiback} and spin 1/2 particle \cite{Majhifermi}  where it has been shown that the coefficient $\beta_1$ is related to trace anomaly. The only difference is the factor before the integration. This agrees well with the earlier conclusion \cite{Duff,Fursaev} where using conformal field theory technique, it was shown that $\beta_1$ is related to trace anomaly and is given by,
\begin{eqnarray}
\beta_1=-\frac{1}{360\pi}\Big(-N_0-\frac{7}{4}N_{\frac{1}{2}}+13N_1+\frac{233}{4}N_{\frac{3}{2}}-212N_2\Big).
\label{coff10}
\end{eqnarray}
Here `$N_s$' denotes the number of fields with spin `$s$'. For gauge field case $N_{1}=1$ and $N_0=N_{\frac{1}{2}}=N_{\frac{3}{2}}=N_2=0$ whereas for gravitino case $N_{\frac{3}{2}}=1$ and $N_0=N_{\frac{1}{2}}=N_{1}=N_2=0$.

\section{Conclusions} 
We have shown that photon and gravitino can tunnel through the event horizon of a black hole just like spin zero and spin half particles. Thus our present work is a natural extension of the Hamilton--Jacobi method previously developed in \cite{Majhifermi,Majhibeyond,Majhiback,Debraj}. In case of photon tunneling, presence of gauge freedom makes the analysis more complicated than the studies for other particles. Nevertheless we have successfully employed the Hamilton--Jacobi method to compute the semiclassical single particle action, and from that, the tunneling amplitude of photon. This has been done by following two different methods. In the first method, we started from a gauge fixed action and calculated the equations of motion for the photon field in a general curved spacetime background. Using the Hamilton--Jacobi ansatz in this equation of motion we obtained the single particle action and tunneling amplitude. After that, principle of detailed balance has been used to recover the semiclassical Hawking temperature. In the other method, starting from the standard Maxwell action in a curved geometry, we follow the previous analysis to obtain a differential equation of the single particle action. Only at this point we used the gauge freedom of photon by considering the Lorentz gauge condition. This, under semiclassical approximation, gave another differential equation. Combination of these two equations produce the same solution of the action. This naturally gave the same semiclasical black hole temperature.

In this paper we have also improved the semiclassical results by incorporating first order quantum effects in the theory. For that we generalized the second method by taking into account the $\hbar$ order equations which come from the Lorentz gauge condition and the Maxwell equation in a gravitational background. Interestingly, it has been found that the correction term of the single particle action is proportional to the semiclassical contribution -- exactly as happens for the scalar and Dirac particles. By dimensional argument, the proportionality constant was shown to be related with the mass of black hole. The corrected action eventually led to the modified Hawking temperature which is in complete agreement with the result obtained earlier\cite{Majhifermi,Majhibeyond}. In our knowledge, existing analysis of tunneling formalism involved emission of spin zero, spin $\frac{1}{2}$ or spin $\frac{3}{2}$ particles from black hole, without discussing the tunneling of photon. In that sense our work fills an important gap present in the literature.

The formalism was applied equally well to the gravitino tunneling case. Previously this was studied \cite{Mann} only upto semiclassical level. Here we have incorporated all the quantum corrections to get the modified black hole temperature and the Bekenstein--Hawking area law. Expectedly, the area law involved logarithmic area correction together with the standard inverse power of area term. Finally, the coefficients of the logarithmic term of entropy which was related with trace anomaly were calculated for both photon and gravitino. This completed our analysis.

\end{document}